# Web Services Non-Functional Classification to Enhance Discovery Speed

Mamoun Mohamad Jamous[1], Safaai Bin Deris[2]

Department of Software Engineering, Faculty of Computer Science and Information Technologies

Universiti Teknologi Malaysia, Skudai, 81310, Johor Bahru, Malaysia

[1]hjmamoun2@live.utm.my

[2]safaai@fsksm.utm.my

**Abstract**
Recently, the use and deployment of web services has dramatically increased. This is due to the easiness, interoperability, and flexibility that web services offer to the software systems, which other software structures don't support or support poorly. Web services discovery became more important and research conducted in this area became more critical. With the increasing number of published and publically available web services, speed in web service discovery process is becoming an issue which cannot be neglected. This paper proposes a generic non-functional based web services classification algorithm. Classification algorithm depends on information supplied by web service provider at the registration time. Authors have proved mathematically and experimentally the usefulness and efficiency of proposed algorithm.

***Keywords:*** *Web services, web service discovery, web service classification.*

## 1. Introduction

Finding suitable web services for the end user or a service oriented system developer requires skills until now, and takes remarkable time even the number of web services is not very big. classification of web services is one methodology that can be used in order to enhance the speed of web service discovery process. Until now, and after the termination of the UDDI project [1], the biggest web services registry to the best of authors knowledge is Seekda [2] with 28 thousand registered web services. Another web service registry is BioCatalogue [3] with two thousand registered web services specialized in bioinformatics. Newly beta-release Service-Finder [4] claims to have 25 thousand registered web services, however most queries produced handing error while authors were testing its capabilities. In both first two portals, results takes remarkable time to load. Moreover, many of the returned results are not related to the user need. For example, if a user is looking for a free weather forecasting web service, Seekda will retrieve web services with term "free" in the description without confirming that this web service is actually free to use or not. Moreover, the user need to read descriptions of retrieved web services in order to confirm if it is free or commercial. Authors are trying with this work to provide discovery processes with a classification solution to enhance discovery speed, and to help retrieve only classes of web services that have been selected by user during the search.

## 2. Related Work

Classification of web services is the act of grouping similar web services into groups. The similarity among a group of web services depends on different criteria, which leads to different classification methods. Classification enhances the speed of web service discovery process. Moreover, classification of web services increases the accuracy of discovering the right service for the specified need. Web services can be classified in different criteria. The following are the criteria being used in classification of web services in recent publications.

2.1. Behaviour

Classifying web services by the functionalities it provide. For example, informative web services is a category were web services provide information without required input from the consumer. For example, weather forecasting, money exchange rates, and global time services. Hongbing in [5] proposes an automated classification algorithm which depends on functional features such as inputs and outputs extracted from web services description files. Hongbing classification criteria depends on a standard taxonomy by UNSPSC.

2.2. Ontology and semantics

Crasso [6] proposed ontological based classification of web services inspired by vector space model. Each word from a web service description file has a weighting scheme based on TF-ITF scheme. His work shows flexibility in managing web services description, even though sources of words representing web services was not precisely determined. One defect of ontology based classification is the fact that different ontologies may cause different classification for the same web service, which may lead to ambiguity in web services discovery. Kehagias [7] proposed a semantic based classification of WSDL files through 3 layers of categorization. His work deployed WordNet [8] as ontology references and PorterStemmer [9] for unique keywords extraction from WSDL description files.

### 2.3. Context Domain

Context domain classification has shown good results in clustering and grouping of web services. This is because consumers normally look for a web service by searching its domain. Abujarour et al [10] proposed an automatic classification algorithm which retrieve crawled web service description files from the web, stem their description and tags, then hash features of each web service using SimHash [11] function and classify web services depending on a domain classification extracted from programmableweb.com web page.

### 2.4. Quality of service (QoS)

Lee et al. proposed at [12] a web service quality management system WSQMS, which they have integrated with UDDI registry *tModel* component in order to link each web service with its quality of service parameters, and then used deployed these parameters in classifying web services. The problem of qos categorization is the lack of semantics, which is highly needed by discovery algorithm. However, QoS categorization is very helpful for web services clustering and filtering, which highly helps end user on making decision of what web service to choose among a group of similar functionality web services.

## 3. Proposed Classification Algorithm

Web services will be classified and stored into classes according to a non-functional criteria. These classes belongs to different criterions, which are predefined and provided by the web service registry. Fig 1 illustrates two non-functional criteria used in our proposed model. Classification attributes values are provided by web service provider during registration of the web service to the registry. However, for the experiment authors are building, and the examples illustrated later, classification of web services are generated randomly for each web service.

### 3.1. Definitions

This section will contain definitions of classification terms used in the classification criteria in this work:
- Free Unlimited: a web service is totally free and can be used for unlimited times.
- Free Limited: the web service is free for a limited time or number of usage times (trial).
- Subscription: the web service can be used only if the user is a subscribed customer, which means he pays some fee for some period of time to use the service, such as accessing a commercial magazine.
- Pay-per-use: the web service performs its functionality every time the customer pays for it, such as ordering flowers or buying software online.
- Collective: the web service has no output. Means that it does not respond to the end user with any information rather than a notification that the invocation was successful. An example of this type of web services is reporting web services.
- Notifying: in this case, the web service does not require any input from the end user in order to function. It only publishes information, such as weather forecasting and money exchange rates.
- Interactive: the web service receives input and produces output while interacting with the end user. An example of this category of web services is a credit card verification service, or a ticket booking service.

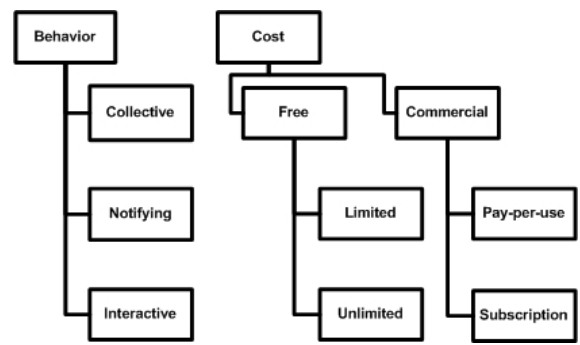

Fig. 1: non-functional classification criteria levels.

Our proposed classification model is meant to be supplied by web service provider by selecting the classification criteria that suits their web service while they are registering it at the web service registry. An algorithm will translate the web service provider selections into an id. This id will represent the non-functional classes of the web service. The algorithm gives each classification level two digits to be represented. Since the proposed classification has three levels, classification id will consist of 6 digits. Fig. 2 illustrates an example of a classification id generation.

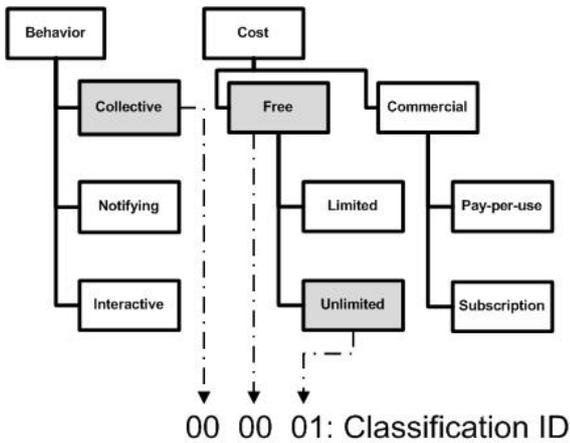

Fig. 2: classification ID generation example.

Considering three levels of classification, the highest level is behaviour-based, the second and third levels cost-based. Fig. 2 shows an example of generating a classification ID for a totally free and collective web service.

## 4. Experiment

In order to proof efficiency of proposed algorithm empirically, a prototype named WSDis was developed, which implements classification and discovery algorithms with QoS filtering model. Discovery algorithm and QoS filtering model are not reported in this paper.

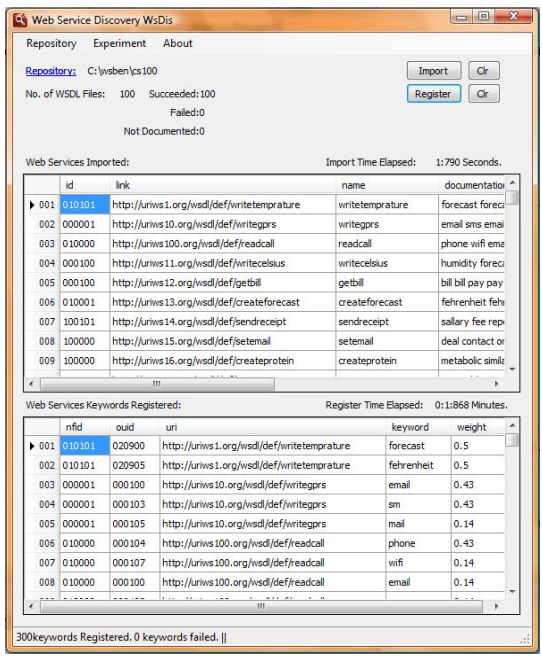

Fig. 3 WSDis main form "Web Services Registration"

WSDis is developed in C# and it has two main stages, (1)Registration: were web service description files are categorised and indexed, the second stage is (2)Discovery: where a GUI is provided for searching the web services registered in the earlier stage.

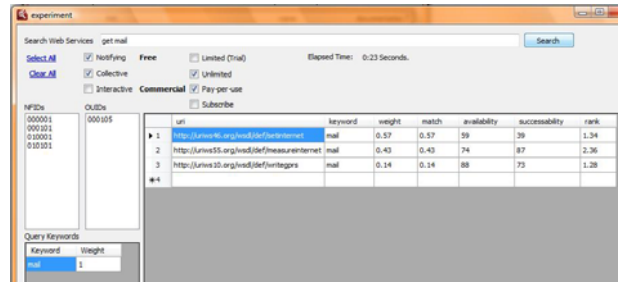

Fig. 4 WSDis experiment form "Web Service Discovery"

Since experiment targets enhancement of speed, web services are distributed randomly under the classification criteria during registration stage. classification criteria was described earlier in this paper. Fig. 3 shows the main form of WSDis where web services description documents are imported and registered , and Fig. 4 shows search form where discovery stage takes place assisted with selection process, where classification algorithm functions.

### 4.1. Web Services Distribution and Coverage

This expression "web services distribution" implies the arrangement of web services in categories under the specified classification criterion.

The expression "web services coverage" implies the web services which are included in the discovery process which are included in the categories of classification criteria.

It is important to show the distribution of web services in order to relate the amount of time saving with the percentage of web services covered by selection process. Table 1 lists the classification distribution for the five datasets used in our experiment.

Table 1: Distribution of web services in classification criteria

| Dataset Size: | 100 | 500 | 1000 | 5000 | 10000 |
|---|---|---|---|---|---|
| Collective | 23% | 31% | 33% | 34% | 33% |
| Notifying | 42% | 35% | 32% | 31% | 33% |
| Interactive | 35% | 32% | 33% | 33% | 33% |
| Free Limited | 35% | 26% | 23% | 25% | 24% |
| Free Unlimited | 31% | 25% | 26% | 24% | 25% |
| Pay-per-Use | 11% | 24% | 24% | 25% | 24% |
| Subscribe | 23% | 23% | 25% | 24% | 25% |

Size of dataset represents the number of WSDL files used. Classification criteria are Behaviour-based (Collective,

Notifying, and Interactive) and Cost-based (Free limited, Free unlimited, Pay-per-use, and Subscribe). Numbers in Table 1 shows distribution in each criterion separated from the other.

4.2. Implementation

For each dataset, web services are imported and registered in the WSDis prototype, then a query was conducted 4 times:
- With one category selected.
- With two categories selected.
- With six categories selected.
- With all categories selected: means that the selection algorithm is not used and all web services are covered.

The same query were used in all runs over all datasets in order to make the comparison valid. For each run, the prototype records the time elapsed, which we compare each time with the 4th run result (all categories selected) in order to calculate the percentage of time saving. The following section discusses the results and compares experimental results with mathematical calculation.

## 5. Results and Discussion

Since our proposed classification is generic, authors find it necessary to prove its efficiency and usefulness mathematically. Then followed by an experiment conducted on an ontology-based discovery algorithm proposed by authors of this paper.

4.1. Mathematical Proof

In this section, authors will provide a mathematical proof for what it was claimed earlier, that the non-functional classification of web services will enhance discovery speed.

**Hypothesis**: non-functional classification of web services enhances discovery speed.

**Proof**: during search for a match to a value of a non-functional id of a web service record, if a match took place, a further processes should take place, otherwise, the matching process should continue and check the next record.

Let's say that the time required for checking a match/non-match status for a non-functional id is $t_{nf}$, and the time required for the discovery matching process to check a record $t_{disc}$.

This means that for each record, the time spent is either $t_{nf}$ or $t_{nf} + t_{disc}$. Let's say that number of matched records is x and non matched records is y. Total time used to run through all records and find matching records can be calculated in the following formula:

$$\text{Time} = x*(t_{nf} + T_{disc}) + y*(t_{nf}) \quad (1)$$

Time spent for discovery matching process is definitely bigger than time used for non-functional matching. This is due to the reasoning nature of matching and the involvement of several components in discovery algorithms, especially ontology reasoning or novel proposed additional web service descriptions like semantic descriptions. However, we consider the simplest discovery matching process may require time at least equal to the non-functional matching process. Having this consideration in mind, we can simplify the previous formula making $t_{nf} \approx t_{disc} = t$.

$$\text{Time} = x*(2t) + y*(t) \quad (2)$$

Now, let us calculate the time needed to check matching of both non-functional and discovery matching processes. For each record, there will be $t_{nf}$ and $t_{disc}$. Which means:

$$\text{Time} = (x+y)*(2t) \quad (3)$$

Time saving by deducting Eq. (2) from Eq. (3) can be calculated as in Eq. (4) as follows:

$$\text{Time saving} = (2xt + 2yt) - (2xt + yt) = yt \quad (4)$$

Since y > 0, and t > 0, thus, time saving > 0. In case y = 0, it means that user decided not to user non-functional classification during discovery, which yields the system will go through all records matching non-functional and ontology. It is obvious that there will be no time saving. Percentage of time saving out of total time can be calculated by Eq. (5). TSP is the abbreviation for Time Saving Percentage as.

$$\text{TSP} = \frac{\text{time saving}}{\text{whole time}} = \frac{Yt}{2xt+yt} = \frac{Y}{2x+y} \quad (5)$$

Fig. 5 shows the relation between web service classification coverage and time saving. Classification coverage means the number of web services included in the discovery process using selected classification criteria by the user.

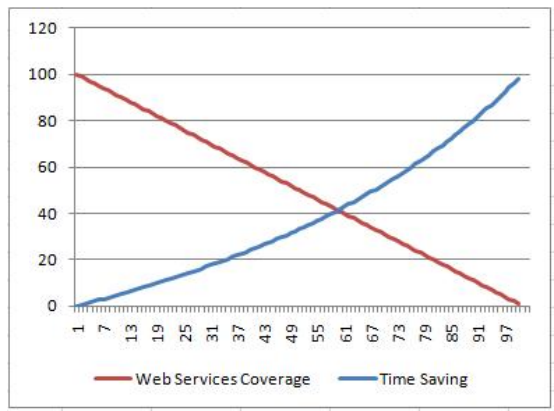

Fig. 5 Relationship of non-functional classification coverage and time saving.

Coverage is inversely proportional with time saving as Fig. 5 illustrates. However, experimental results depends on the distribution of web services among the categories, and real relation should vary slightly from Fig. 5. as Table 1 shows. In the next section, experimental results will be exposed.

4.2. Experimental Results

Five datasets of web services were generated using a customized version of WSBen [13] framework. Authors have made the customization to WSBen in order to generate web services with meaningful name and documentation. dataset 1 contains 100 web service WSDL description files, dataset 2 contains 500, dataset 3 contains 1000, dataset 4 contains 5000, and dataset 5 contains 10000 WSDL files. Each dataset was deployed four times, each time with different classification coverage. First time with 1 category covered, second time with 2, third time with 6, and the last time with all categories included (this is equal to not using classification at all for discovery). Fig. 6 illustrates time saving in the 5 different datasets mentioned earlier.

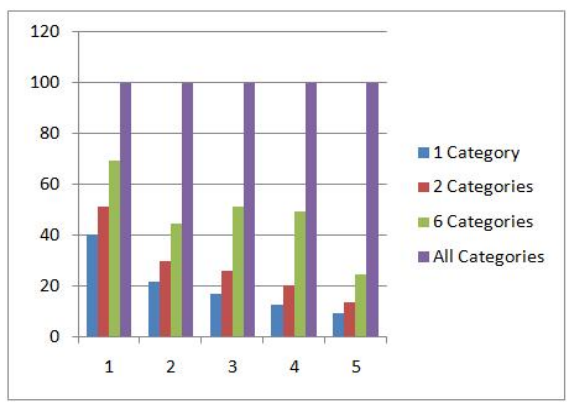

Fig. 6: time saving for different non-functional classification coverage.

Results have shown remarkable saving of time - 50% to 90% - when using one or two categories, especially when the number of web services is big, such as in dataset 5 of 10000 web service description files. this is due bypassing the reasoning logic of matching, which consumes nontrivial time in most cases of discovery algorithms.

## 6. Conclusion

Number of web services that being publically available is increasing tremendously recently, and the discovery process speed issue is coming to the surface. Classification of web services adds remarkable benefits to web service discovery processes. It helps categorize web services and contributes to the discovery process speed enhancement. Our proposed non-functional classification depends on three layers of classification criteria. Mathematical and experimental proofs has been conducted, and they have validated that our proposed classification is useful and efficient.

**Mamoun Mohamad Jamous** received his Bachelor degree in computer science from Sudan University for Science and Technology at 2005. Msc in Computer Science at Universiti Teknologi Malaysia at 2008. He is currently a PhD student at Universiti Teknologi Malaysia. His research interest is Web Services Discovery, Software Architecture, Software Requirements, and Computer Architecture.

**Safaai Bin Deris** is a full professor of Computer Science at Universiti Teknologi Malaysia. His research interest is Software Engineering, Service-Oriented Architecture, Bioinformatics, and Scheduling, and Artificial Intelligence computing.